\documentclass[preprints,review,accept,moreauthors,pdftex]{Definitions/mdpi} 

\firstpage{1} 
\pubvolume{1}
\issuenum{1}
\articlenumber{0}
\pubyear{2022}
\copyrightyear{2022}
\datereceived{25 January 2022} 
\dateaccepted{} 
\datepublished{} 
\hreflink{https://doi.org/}

\newcommand{\egret}{\textit{EGRET}}
\usepackage{amsmath}
\let\oldAA\AA
\renewcommand{\AA}{\text{\normalfont\oldAA}}

\usepackage{natbib,paracol}

\Title{The Blazar sequence and its Physical Understanding}

\TitleCitation{The Blazar Sequence and its Physical Understanding}

\Author{Elisa Prandini $^{1}$\orcidA{} and Gabriele Ghisellini $^{2}$\orcidB{}}

\AuthorNames{Elisa Prandini and Gabriele Ghisellini}

\AuthorCitation{Prandini, E.; Ghisellini, G.}

\address{%
$^{1}$ \quad Department of Physics and Astronomy, University of Padova, Via Marzolo 8, 35131 Padova (PD) - Italy\\
$^{2}$ \quad INAF Osservatorio Astronomico di Brera, via E. Bianchi 46,
23807 Merate (Lc) -  Italy}

\corres{Correspondence: elisa.prandini@unipd.it; gabriele.ghisellini@inaf.it.}

\abstract{Introduced in 1998 to attempt a first unified view of the broad-band emission properties of blazars, the blazar sequence has been extensively used in the past 25 years to guide observations as well as physical interpretation of the overall emission from these  galaxies. In this review, we describe the evolution of the sequence along with the tremendous advances in the observational field, in particular in the gamma-ray band. A new version of the sequence built on TeV-detected objects is also presented. Two extreme classes of objects (MeV and hard-TeV blazars) are included in the discussion, given their relevance for future observatories. Finally, the current physical understanding at the base of the sequence is presented along with the major criticisms to the blazar sequence.
}

\keyword{blazars; astrophysics; TeV astronomy} 

\begin{document}

\section{Introduction: the blazar paradigm}

In retrospect, we can identify the discovery of the first quasar, 3C 273 \citep{1963Natur.197.1040S}, as also the discovery of the first blazar. 3C~273 is in fact a variable radio source with an optical counterpart whose detailed spectroscopic analysis revealed an extragalactic nature ($z = 0.158$). 
It was 1963, and tremendous advancements in the multiwavelength and, more recently, multimessenger observations of the Universe have been done since then. 
Almost sixty years after this first identification, thousands of blazars now populate the extragalactic sky. Blazars are by far the most numerous class of persistent extragalactic sources in the gamma-ray band \citep[see e.g.,][]{2020ApJS..247...33A}. A blazar is also the first extragalactic source associated as a likely counterpart for high-energy neutrino emission \citep{2018Sci...361.1378I}. 

In the past decades, the increase of the number and quality of observations from the radio to the gamma-ray band along with improved theoretical models and numerical simulations allowed to establish a paradigm explaining the blazar and in general the AGN structure and many of the observed properties \citep{1993ARA&A..31..473A,1995PASP..107..803U,2014ARA&A..52..589H}. 
Still, a number of unknowns remain on the AGN phenomenon. The ones most relevant to the blazar case are: what is the physical driver of the relativistic jet and how it connects with the BH mass and the accretion rate; what is the rate of cosmic rays and neutrinos emitted in a blazar's jet and how does it connect with the gamma-ray emission mechanisms; what is the origin of the observed variability and how it is connected with the overall multi-wavelength and multi-messenger emission \citep[see e.g.,][for a recent  review]{2017A&ARv..25....2P}.

Blazars are jetted active galactic nuclei (AGNs) with a jet pointing at a small angle ($<10^\circ$) toward the observer. 
Roughly 10\% of jetted-AGNs are blazars \citep{2017A&ARv..25....2P}. Blazars come in two flavours:
Flat Spectrum Radio Quasars (FSRQs) and BL Lac objects, depending on the width of their emission lines. FSRQs exhibit lines with an equivalent width $>5$\,\AA. In BL Lac objects, the lines are fainter or even absent.

\begin{figure}
    \centering
    \includegraphics[width=0.35\textwidth]{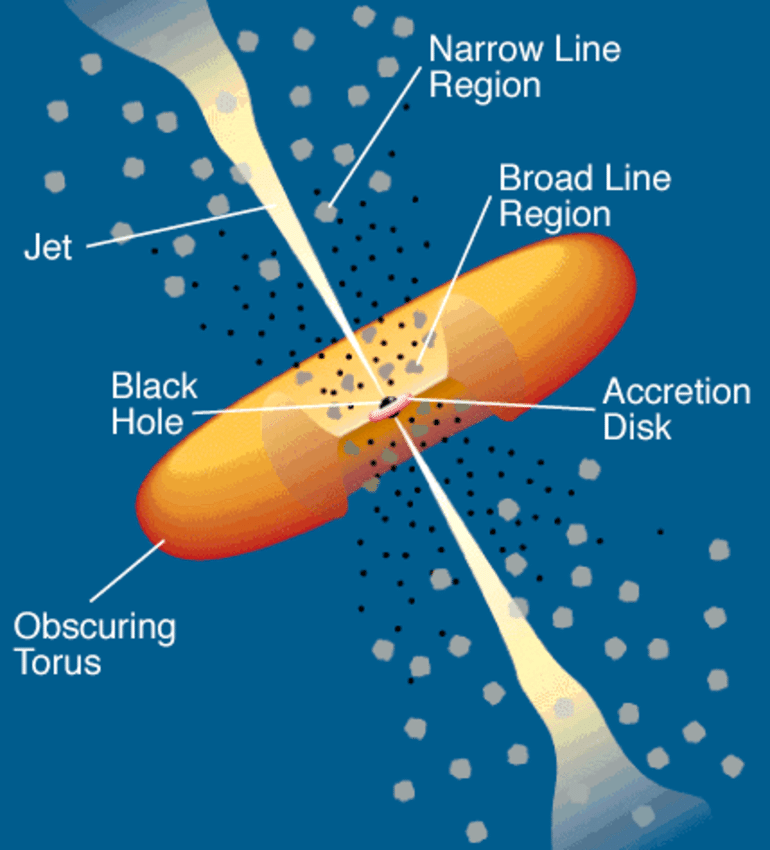}
    \caption{Jetted AGN simplified scheme from \citet{1995PASP..107..803U}.}
    \label{fig:agn_structure}
\end{figure}

The current paradigm, see the scheme in Fig.~\ref{fig:agn_structure}, foresees that blazars, and more in general jetted-AGNs, are composed of:
\begin{itemize}
    \item A supermassive black hole with a mass between 10$^{7}$ to 10$^{10}$ $M_{\odot}$, located in the center of an elliptical galaxy.
    \item A flow of mass feeding the black hole, named accretion disk.
    \item Two highly collimated jets arising from the proximity of the central object and extending to several kpcs. Within the jet, there are regions where particles are accelerated to ultra-relativistic energies. 
    \item In FSRQs, an obscuring torus of dust that surrounds the accretion disc.
\end{itemize}

In FSRQs, the analysis of optical-to- X-ray data has shown that the accretion occurs through a thin and hot accretion disc. This disc is surrounded by a hot thermal corona and by two regions filled with clouds of gas, a narrow line region (NLR) and a broad line region (BLR). 
In BL Lac objects, the absence of broad emission lines and of any sign of thermal emission produced by the accretion disk suggests that the two properties are related. 
In other words, the weakness of the accretion disk radiation may be due to the accretion occurring in the Advection Dominated Accretion Flow (ADAF) regime \citep[see e.g.,][]{2000ApJ...539..798N,1999MNRAS.303L...1B,1997ApJ...478L..79N}. The ADAF regime is radiatively inefficient and characterized by a dramatic decrease of ionizing UV photons, that are therefore unable to photo--ionize the clouds of gas responsible for the broad lines \citep[see e.g.,][]{1997ApJ...477..585M}. This coarse division into BL Lac objects and FSRQs, however, is in some cases arbitrary as transitional objects displaying mixed properties exist \citep[see, e.g.,][]{2018A&A...617A..30M}.

\begin{figure}{}
    \centering
    \includegraphics[width=0.35\textwidth]{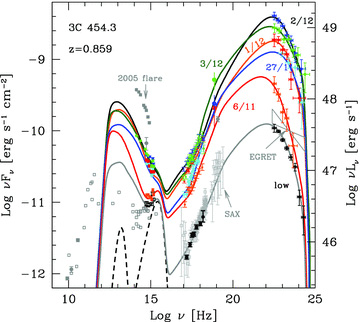}
    \includegraphics[width=0.35\textwidth]{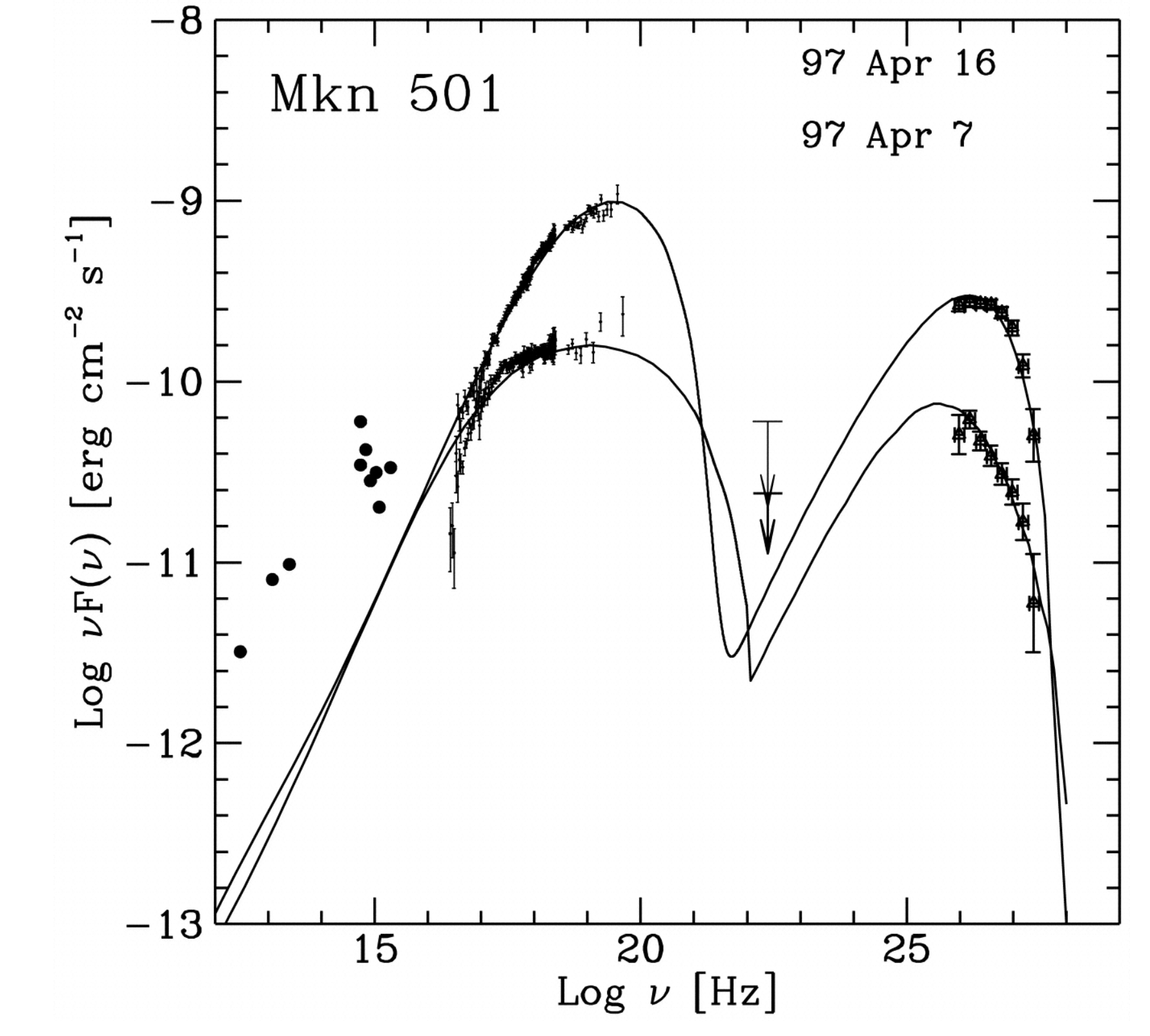}
    \caption{Examples of spectral energy distributions (SEDs) of a FSRQ (right, 3C~454.3) and a BL Lac object (left, Mkn~501) from multi-epoch observations. In the SED of 3C~454.3, in addition to the jet emission, we can identify an accretion disc component, the X-ray corona contribution and the IR emission from the torus (dashed black lines), from \citet{2011MNRAS.410..368B}. In Mkn~501, instead, only the contributions from the jet and host galaxy are evident, from \citet{2001ApJ...554..725T},  \copyright AAS. Reproduced with permission.} 
    \label{fig:fsrq_bllac}
\end{figure}

From the observational point of view, blazar emission covers more than 15 orders of magnitude in frequency, from radio to $\gamma$-rays. 
Blazars are in general highly variable objects, with variations ranging from months to minute timescales \citep{2019Galax...7...28R}. The complexity of the spectral energy distribution (SED) depends on the class of objects considered, with FSRQs showing in general a more articulated SED, with several superimposed contributions. As an example, the multi-epoch SEDs of 2 blazars, the FSRQ 3C~454.3 and the BL~Lac object Mkn~501, are represented in Figure~\ref{fig:fsrq_bllac}.  The SED of 3C~454.3 is interpreted as the superposition of the jet emission (two broad bumps peaking at optical and gamma-ray energies), the IR torus, and the hot corona. Mkn~501 SED is instead fully characterized by the jet emission and a contribution of the host galaxy in the sub-optical spectral domain.

As usual in astronomy, classification based on experimental evidences helps identifying key features and pinpointing the physical mechanisms responsible for the broadband emission registered. Blazars make no exception to this rule. This review is devoted to the \textit{blazar sequence}, a phenomenological sequence proposed in \citet{1998MNRAS.299..433F} to describe the average properties of the SED of blazars. Since this first article, great advancements in the observations, in particular in the $\gamma$-ray band, have questioned and tested the sequence, which was updated accordingly. 
\S~\ref{sec:two} is devoted to an overview of the up-to-date observational properties of blazars complemented by a description of the state-of-the-art emission models. In \S~\ref{sec:three}, we present the blazar sequence and its evolution along the years. We will also propose a new sequence based on TeV data. In \S~\ref{sec:four} we discuss the extremes of the sequence, namely, MeV blazars and extreme blazars. In \S~\ref{sec:five}, we  outline the physical interpretation of the sequence, while \S~\ref{sec:six} is devoted to the main criticisms against the sequence. 
A discussion on future observations concludes this manuscript in \S~\ref{sec:seven}.

\section{Blazar Emission}\label{sec:two}
Given their broadband and variable SED, coordinated, multi-wavelength observations are the key to accessing a reliable characterization of blazar emissions. 
From the observational point of view, the multi-wavelength approach became accessible only in the '90ies, when \textit{COMPTEL} and \egret~in gamma rays \citep{1993ApJS...86..657S,1993ApJS...86..629T} flanked several X-ray satellites \citep[see][for a review]{2009ExA....25..143G} and opened the high-energy window. A few years later this window was extended to the very high energy domain (VHE, E>100\,GeV) by ground-based Cherenkov telescopes \citep[see][for a review]{universe7110432}. And starting from 2008, the \textit{Fermi} satellite \citep{2009ApJ...697.1071A}, together with the second generation of ground-based Cherenkov telescopes (H.E.S.S. \citep{2003APh....20..111B}, MAGIC \citep{2016APh....72...61A}, VERITAS \citep{2006APh....25..391H})  have literally revolutionized our view and knowledge of the $\gamma$-ray universe.

One of the challenges that the blazar, and in general the high-energy community is facing is that of collecting simultaneous, possibly strictly contemporaneous data in a broad range of frequencies. This often requires a daily-scale human intervention (for target selection, request of observation, and coordination). An effort in this sense, focussed on transient phenomena, is the Astrophysical Multimessenger Observatory Network (AMON) \citep{2013APh....45...56S} and, under development, a number of broker alert systems designed for handling the Vera C. Rubin Observatory Legacy Survey of Space and Time alert stream, such as the one discussed in \citet{2021MNRAS.501.3272M}.



\subsection{Emission Models}
The most popular models to interpret the blazar emission differ in particles and are: 
\begin{itemize}
\item[-]\textit{One zone, leptonic} \citep{1985A&A...146..204G}, that considers that most (but not all!) luminosity is produced in a well-defined zone at some distance $R_{\rm diss}$ from the central engine by relativistic electrons; 
\item[-]\textit{One zone, hadronic} \citep[e.g.,][]{2013ApJ...768...54B}, that assumes that the relativistic protons are responsible for the emission, even if not directly (except for the proton--synchrotron model). Proton--proton collisions or more likely, photo-hadronic interactions can produce electron positron relativistic pairs that can then radiate;
\item[-]\textit{Multi zone, either leptonic or hadronic} \citep[e.g.,][]{2013MNRAS.436..304P}, assumes that the particles are accelerated and radiate all along the jet in a more or less continuous way. These models then consider that the density of the emitting particles and the magnetic field are a (power--law) function of the distance from the black hole. In these models, the jet geometry (paraboloidal or conical) plays a crucial role.
\end{itemize}
{\it Radiative processes:} In all cases the main radiation processes of the jet are the synchrotron mechanism for the low energy part and the Inverse Compton (IC)  for the high energy part. If the seed photons for the IC scattering are the synchrotron photons produced by the same electrons producing the high energy Compton component, the process is called Synchrotron--Self--Compton (SSC), while if the main seeds are produced externally to the jet  (disk radiation, BLR lines, torus...) the process is called external Compton (EC). 

{\it Spine--layer structure:} Observational evidence
\citep[e.g.,][]{2004ApJ...600..127G} and important $\gamma$--ray emission from misaligned sources \citep[e.g.,][]{2012ApJ...751L...3G} led to the suggestion that the blazar jet is structured. \citet{2005A&A...432..401G} suggest that a high velocity spine is surrounded by a slower layer ($\Gamma\sim 13$ and $\Gamma\sim 3$, respectively). The spine sees the layer radiation beamed and this enhances its Compton emission. On the other hand, also the layer sees the radiation from the spine as beamed, and thus also the Compton emission of the layer is enhanced. 

{\it Thermal emission: BL Lacs.} In BL Lacs we see no sign of thermal emission, leading to the suggestion that: 
\begin{enumerate}
\item The accretion regime is not radiatively efficient;
\item This corresponds to a paucity of ionizing radiation, corresponding to the absence of broad emission lines;
\item There is no molecular torus, as first suggested by \citet{1999A&A...349...77C};
\item All these properties can be understood if the accretion luminosity, in units of the Eddington one, is smaller than some critical value ($L_{\rm disk}/L_{\rm Edd}  \le 10^{-2}-10^{-3}$). \end{enumerate}

{\it Thermal emission: FSRQs.} In powerful  FSRQ, instead, we do see directly the accretion disk radiation, besides the broad emission lines and the IR torus component.
This therefore suggests that the high energy emission of FSRQs is likely due to the EC process, while in BL Lacs is SSC (but possibly accounting for the spine--layer structure).

\subsection{Energy budget}
The first duty of the jet is to bring mechanical energy to the extended structures (hot spots and radio lobes).
Radiation instead corresponds to $\sim$10\% of the total jet energy \citep{2014Natur.515..376G}.
It is not easy to calculate the jet power directly from observations, because the radiation is beamed. 
Nevertheless, if we know (i.e., from superluminal motion) the bulk Lorenrz factor $\Gamma$
then we can set a firm lower limit to the jet power, that cannot be smaller than
$L_{obs}/\Gamma^2$ \citep{2014Natur.515..376G}. 
Therefore, assuming 10\% efficiency \citep[e.g.,][]{2012Sci...338.1445N} we have that the total jet power is of the same order, but on average larger, than the accretion disk luminosity \citep{2014Natur.515..376G}. Blazar's jets are therefore the most efficient persistent engines in nature.

\begin{figure}
    \centering
    \includegraphics[width=0.32\textwidth]{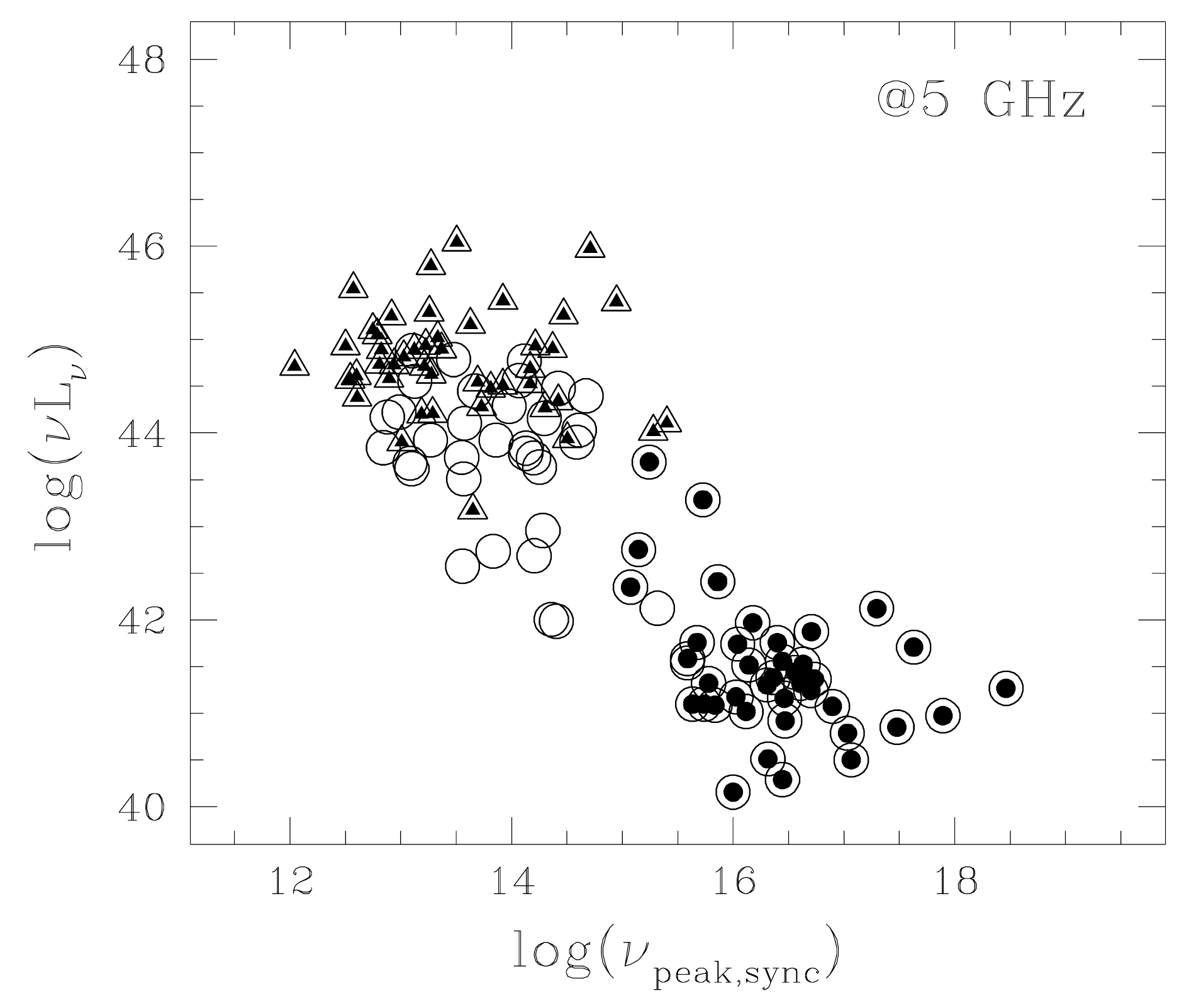}        \includegraphics[width=0.32\textwidth]{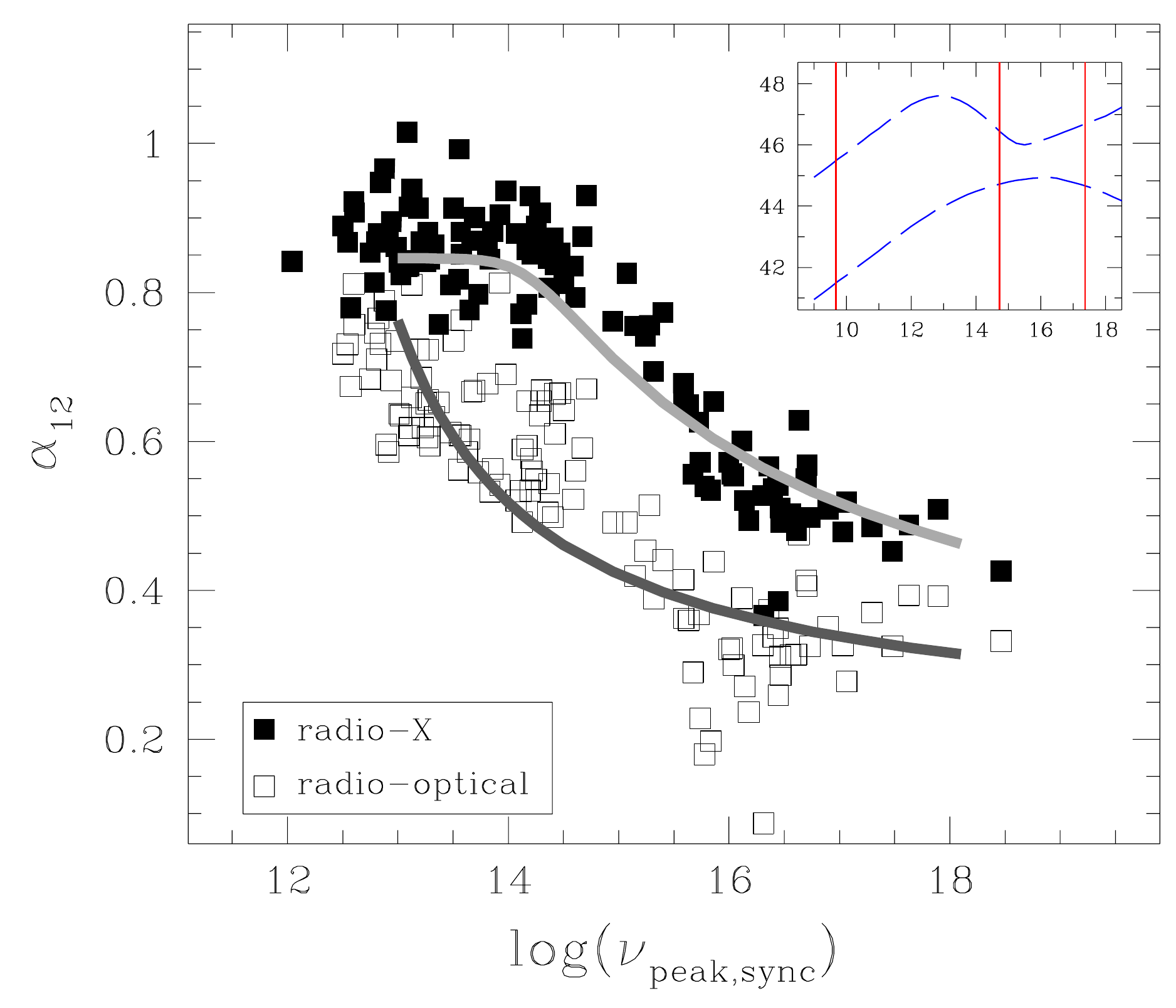}
    \includegraphics[width=0.32\textwidth]{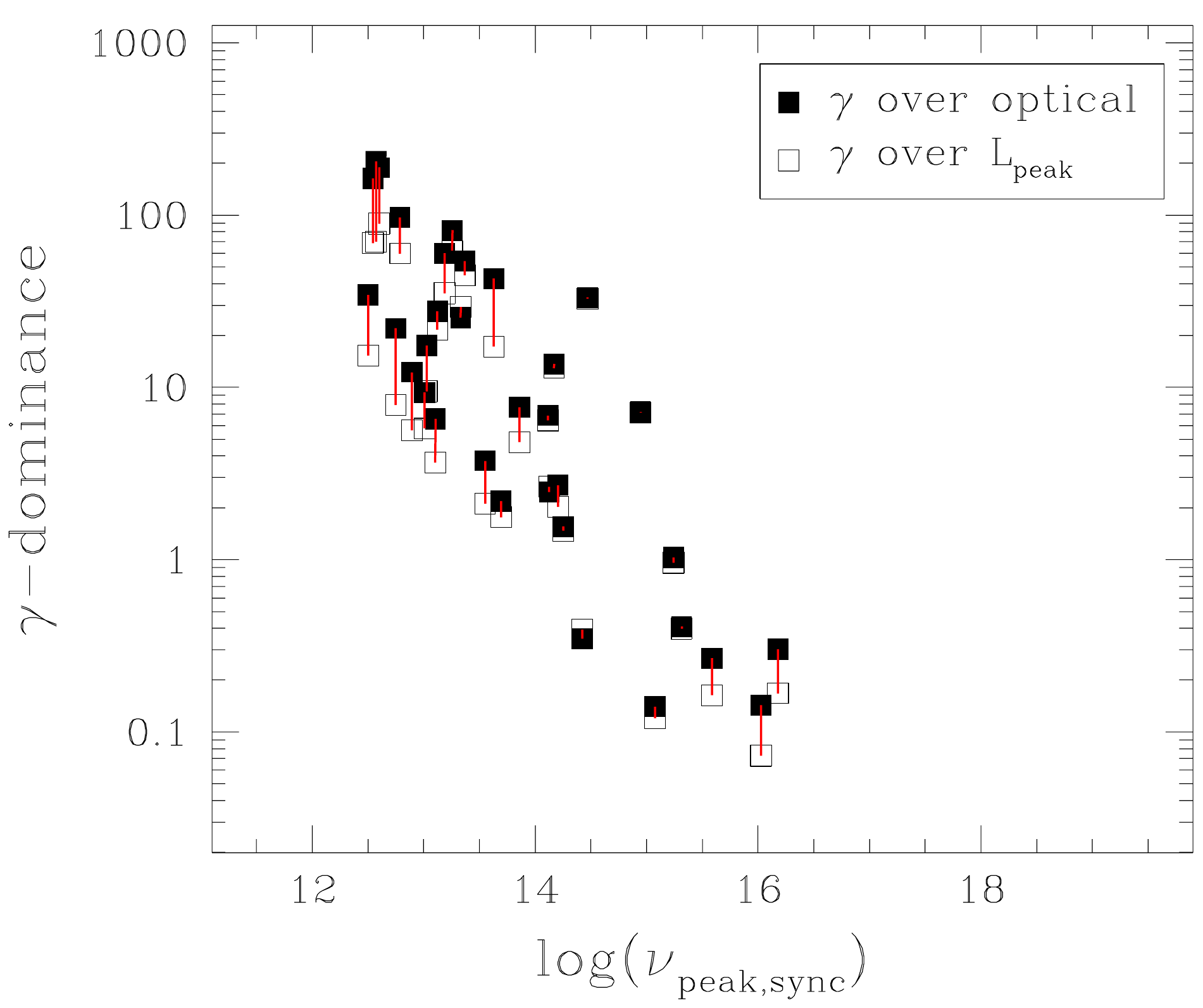}        
    \caption{Dependance of the synchrotron peak frequency with the radio luminosity (left), spectral indices  $\alpha_{RO}$ and $\alpha_{RX}$ (central), and $\gamma$-ray dominance (right). Figures from \citet{1998MNRAS.299..433F}}
    \label{fig:sequence_plots}
\end{figure}
\begin{figure}
\centering
    \includegraphics[width=0.37\textwidth]{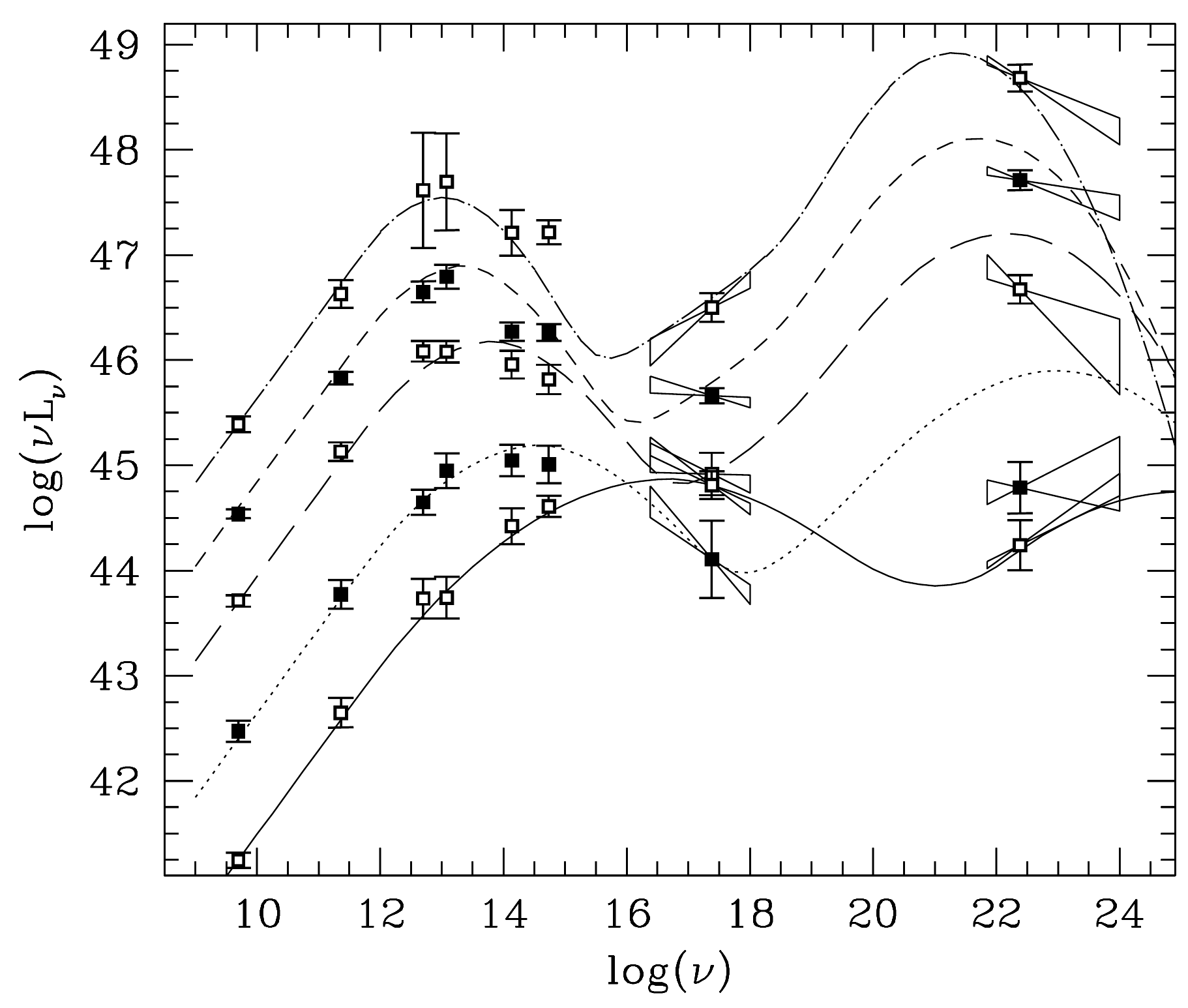}
    \includegraphics[width=0.33\textwidth]{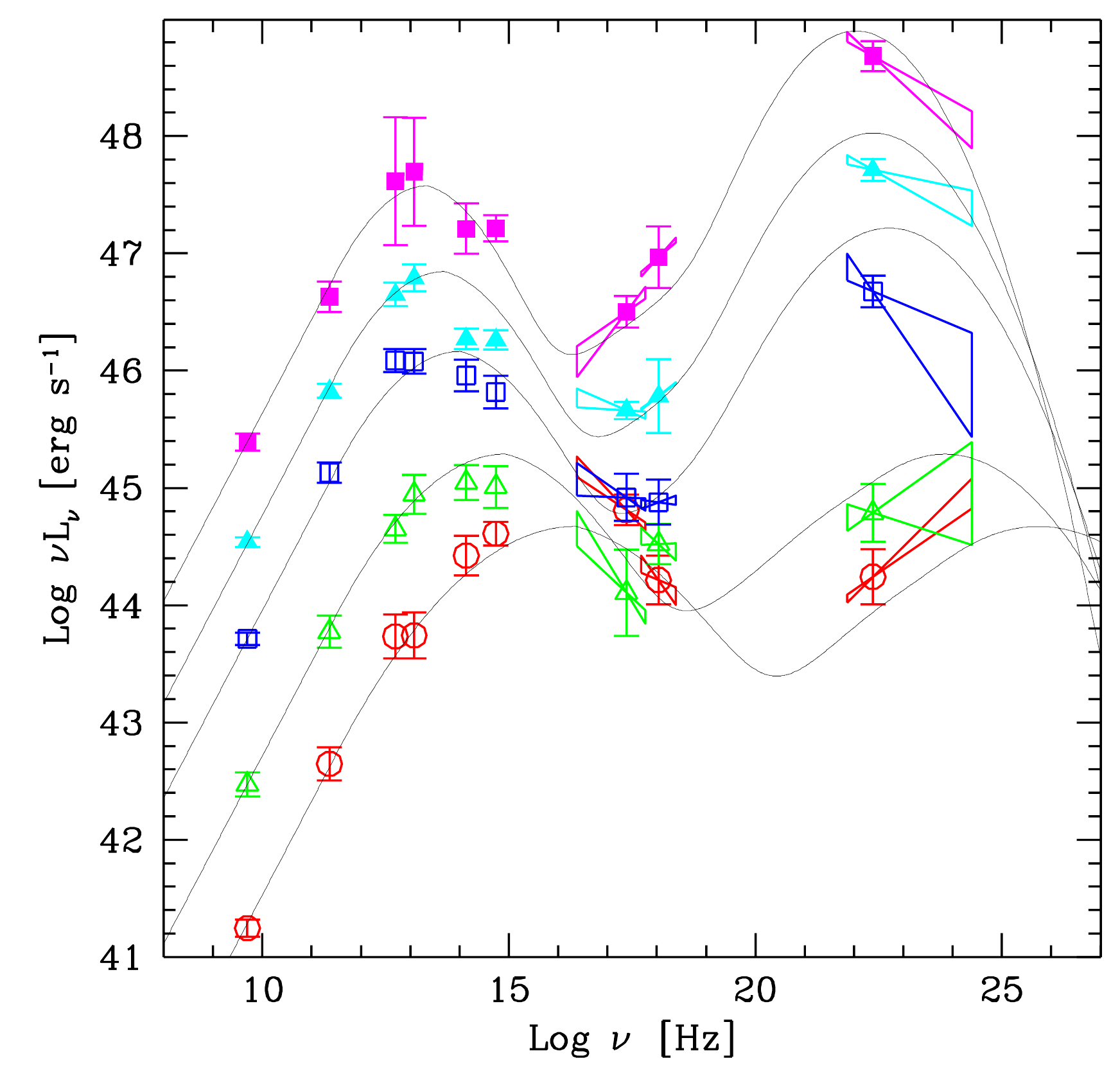}
    \caption{Left: The original blazar sequence, from \citet{1998MNRAS.299..433F}. Right: the updated sequence, from \citet{2001A&A...375..739D} , reproduced with permission \copyright ESO, built with additional [2–10 keV] average spectral indices.}
    \label{fig:sequence_fossati}
\end{figure}

\subsection{Key observational features}
The key features of blazar emission in different bands can be interpreted in light of the above-mentioned emission models as follows:
\begin{description}
\item[Radio:] The radio emission of blazars is dominated by the beamed jet emission, and only at sub-GHz frequency there is the emergence of radiation produced in the extended structures, such as hot spots and lobes. 
The radio spectrum is flat, i.e., $F(\nu) \propto \nu^{-\alpha}$ with 
the spectral index $\alpha$ around zero. This is due to the superposition 
of different jet zones self-absorbing at different frequencies. 
This was thought to be a "cosmic conspiracy" in the old days
\citep{1980ApJ...238L.123C},  but now it can be understood on the 
bases of simple conservation laws: the conservation of emitting particles along the jet demands that their density $n\propto R^{-2}$
(where $R$ is the distance from the start of the jet) while 
conservation of the Poynting flux demands that the magnetic field 
$B\propto R^{-1}$. With these scalings, one derives a flat synchrotron 
radio spectrum and a self--absorption frequency $\nu_{\rm t} 
\propto R^{-1}$. This implies that at smaller frequencies the emitting 
region is larger and the flux is less rapidly variable. 

\item[IR:] In the IR band, we can have the contribution of the jet and, for FSRQs, of the molecular torus. For FSRQs, the sub--mm band is where the synchrotron peaks, and this corresponds to the self--absorption frequency of the innermost emitting region.

\item[Optical:] In the optical band we have the contribution of the jet continuum and, for FSRQs, of the low frequency part of the accretion disk. This component usually dilutes the polarization of the synchrotron component. For low power BL Lacs we also have the contribution of the host galaxy.

\item[UV:] For FSRQs, in the UV band we have the contribution of the steep tail of the synchrotron component and of the accretion disk, that becomes increasingly dominant as the total power increases.
For BL Lacs we only have the synchrotron emission: if the spectrum is rising (in $\nu F_\nu$) we call these objects High frequency peak BL Lacs (HBLs) or "blue" BL Lacs, if the spectrum is decreasing we have a Low frequency peak BL Lac (LBL) or "red" BL Lac. We note, however, that both FSRQ and LBL classes hold transitional objects with mixed properties whose classification is somehow arbitrary.

\item[X-ray:] In FSRQs the X--ray spectrum is increasing (in $\nu F_\nu$) with a spectral index $\alpha$ around 0.5, generally flatter than what expected for a thermal X--ray corona (for which $\alpha\sim 0.7-1$). 
For BL Lacs we can have a rising spectrum for LBLs, while in the case of HBLs the spectrum can peak in the X--ray band. 

\item[Gamma rays:] In the $\gamma$-ray band we have the contribution of only the non--thermal beamed component of the jet. 
In the sub-TeV band 
the FSRQs usually show a steep (i.e. $\alpha>1)$ spectrum, while HBLs can have their high energy peak there, or even at larger energies. In the latter case, we call these objects "extreme" BL Lacs. 
At these energies, the Extra--galactic IR and optical Background Light (EBL) can absorb (through photon--photon collisions producing electron--positron pairs \citep[see e.g.,][]{1992ApJ...390L..49S}) high energy photons, making the observed spectrum decreasing almost exponentially. Since the level of the EBL is still  uncertain \citep{2017A&A...603A..34F}, detailed observations of blazars can help to fix it. 

\item[Neutrinos:] neutrinos are the smoking-gun signature of hadronic interactions in blazars \citep{1991PhRvL..66.2697S}. A single neutrino of energy 2\,PeV detected by the IceCube Observatory has been associated so far with a blazar \citep{2018Sci...361.1378I}, the BL Lac object TXS~0506$+$056. 
\end{description}

\section{The blazar sequence: observational approach}\label{sec:three}
\subsection{The original blazar sequence}
The original blazar sequence was proposed in  \citet{1998MNRAS.299..433F} as an attempt to identify observational properties in the SED of blazars. The sequence was built using multi-wavelength data from 126 sources.
It was the first systematic study of the SED of blazars including $\gamma$-ray data above 30\,MeV from \egret\, \citep{1980ITNS...27..364H}. For the compilation of the sequence, the sources were selected from the following samples:
\begin{itemize}
    \item X-ray selected BL Lacs from the \textit{Einstein} Slew survey \citep{1992ApJS...80..257E}.
    \item Radio-selected BL Lacs from the catalogue of extragalactic sources with $F_{5\,\rm GHz}\geq1$ Jy \citep{ 1981A&AS...45..367K}. 
    For the selection, additional requirements were adopted based on radio flatness ($\alpha_R\leq0.5$), optical brightness ($m_{v}\leq20)$, and the weakness of optical emission lines ($EW_{\lambda}\leq 5 \AA)$.
    \item Radio-selected FSRQs proposed by \citet{1992ApJ...387..449P} from the 2-Jy sample \citep{1985MNRAS.216..173W}.
\end{itemize}
Table~\ref{tab:orig_seq_sample} summarizes the number of sources adopted from each catalog. We have also outlined the corresponding number of $\gamma$-ray detected sources both in the MeV-GeV and in the TeV range. 
Once considered that 4 sources were present in both BL Lac samples and 1 source of the X-ray sample was detected only at TeV, the net number of blazars selected for the study becomes 126.

Once defined the source sample, to build the sequence, the fluxes at seven well-sampled frequencies were collected from the NASA Extragalactic Database (NED\footnote{http://ned.ipac.caltech.edu/}). 
When multiple observations were available, they were averaged 
logarithmically (magnitudes), to account for the large variability. 
The seven frequencies considered for the original blazar sequence are: 
radio at 5 GHz, millimeter at 230 GHz, far-infrared at 60 and 25 $\mu$m, near-infrared ($K$ band) at 2.2 $\mu$m, optical ($V$ band) at 5500 \AA, and soft X-ray at 1 keV. 
Additionally, information on $\gamma$-ray spectral properties were used when available. 
In particular, the \egret~ detections were 32 (12 BL Lacs and 20 FSRQs) 
plus only one source TeV-detected (Mkn~501) belonging to the X-ray selected sample. 
Out of the 32 \egret~ detected sources, 28 spectral determinations 
were accessible at the time of this study. 

\begin{table}[]
    \centering
    \begin{tabular}{c|c|c|c}
        \hline
        \hline
        Catalog & Sample & MeV-GeV  & TeV \\
         & & detected & detected \\
        \hline
        \hline
        X-ray BL Lac & 48 & 7 & 2 \\
        Radio BL Lac & 34 & 9 & 1 \\
        Radio FSRQ   & 50 & 20 & 0 \\
        \hline
        \hline
    \end{tabular}
    \caption{Sample of sources adopted for the original blazar sequence \citep{1998MNRAS.299..433F}. The last two columns highlight the number of sources also detected in $\gamma$-rays, at MeV-GeV and TeV energies, respectively.}
    \label{tab:orig_seq_sample}
\end{table}

Before building the sequence, a number of checks were performed by the authors aimed at inspecting the differences between $\gamma$-ray detected and undetected source samples.
They compared the redshift, luminosity, and broad-band spectral indices. No difference emerged between $\gamma$-ray detected and undetected radio-selected BL Lacs and FSRQs. For the X-ray selected sample, the analysis showed a tendency for $\gamma$-ray loud sources  to have a larger radio luminosity and steeper radio and X-ray spectral indices. Indeed, the limited sensitivity of the \egret~ instrument implies that only the $\gamma$-ray brightest sources were detected. \citet{1998MNRAS.299..433F} underline that the average $\gamma$-ray luminosities reported in the  sequence were necessarily overestimated.
20 years later, the study of the properties of thousands of blazars detected at $\gamma$-rays with the \textit{Fermi} satellite
would have confirmed this limitation (\S~ \ref{sec:fermiseq}).

As a second step toward a comprehensive characterization of the broadband emission from the  sample, the authors determined with a simple fitting procedure in the $\nu L_{\nu}$-$\nu$ plane the synchrotron peak frequency $\nu_{peak, sync}$. They then studied the relation between this frequency and several quantities, namely: luminosity in different bands, optical/radio and X-ray/radio flux ratios $\alpha_{RO}$ and $\alpha_{RX}$, $\gamma$-dominance, i.e., the ratio between the luminosity in $\gamma$-ray and that in correspondence to the synchrotron peak. 
The main results shown in Fig.~\ref{fig:sequence_plots}, representing the seeds of the blazar sequence, are:
\begin{itemize}
    \item[-]The radio luminosity strongly correlates with $\nu_{peak, sync}$. In this study, the authors ruled out a possible bias due to $z$ by performing detailed tests. 
    \item[-]$\nu_{peak, sync}$ correlates with both $\alpha_{RO}$ and $\alpha_{RX}$, meaning that the knowledge of either of the two indices allows a first estimate of $\nu_{peak, sync}$, at least in the range 10$^{14}$--10$^{16}$Hz.
    \item[-]$\nu_{peak, sync}$ also correlates strongly with the $\gamma$-ray dominance. This indicates that sources with a synchrotron peak at small frequencies are brighter $\gamma$-ray emitters. 
\end{itemize}

A clear picture emerged from this study: a smooth transition between different properties of the blazar SED appears when considering the synchrotron peak frequency.
This becomes even more evident when five averaged broadband SEDs are built adopting as bin criterion the radio luminosity. This means that, independently from the original classification of a blazar as FSRQ or radio/X-ray selected BL Lac, the average SEDs were built taking into account only the luminosity of the source at 5\,GHz,  available for all objects. The result is the \textit{blazar sequence}, displayed in Fig.~\ref{fig:sequence_fossati} (left). The radio-bin log-(L$_5\,GHz$) intervals are: <42; 42-43; 43-44; 44-45; >45. A set of lines is superimposed on the data, with the low-frequency part composed by a power law connected to a parabolic branch (synchrotron peak and successive steepening). A second power law connects this first peak to a second peak, approximated with another parabolic branch and built following some basic assumptions on the peak position and luminosities in different frequencies. 

The analysis of the average data and their trends reveals that all 
SEDs present a two-peak structure. 
The frequency of the first peak is anticorrelated with the luminosity and moves from 10$^{13-14}$ Hz for brighter sources, to 10$^{16-17}$ 
Hz for fainter sources. A similar effect emerges in the high-energy peak, 
despite the poor sampling, and suggests that the frequencies of the two 
peaks are correlated. 
Another evidence is that sources with higher frequency peaks have
a smaller $\gamma$-ray dominance (now called Compton dominance).


Three years after the compilation of the original sequence, \citet{2001A&A...375..739D} updated it by adding the [2–10 keV] average spectral indices and fluxes for the same source sample introduced in \citep{1998MNRAS.299..433F}. 
The revised sequence is shown in Fig. \ref{fig:sequence_fossati} (right), where only minor modifications were introduced in the parametrization of the SEDs. These modifications were physically driven and reflected the fact that for low luminosity sources there were growing evidences supporting the SSC scenario for BL Lacs.
The work confirmed, on a statistical basis, that more powerful blazars emit the X–rays by the inverse Compton process, while in less powerful objects the dominant mechanism is the synchrotron. 
Thus, the shift in both peak positions proposed by the original sequence was confirmed.

The emergence of the blazar sequence was of uttermost importance in the experimental, phenomenological, and theoretical panorama. First, it suggested that a classification based on the synchrotron peak position, in particular for BL Lac objects classified into low-, intermediate-, and high-synchrotron peaked sources, must have a physical base, given the smooth evolution observed between the blazar classes. This explanation should also account for the Compton dominance behaviour emerging at the highest energies. Moreover, it became evident that the common observational properties between FSRQ and BL~Lacs call for a common structure and underlying physical mechanisms.

From the experimental perspective, the sequence represented a challenge for the emerging VHE $\gamma$-ray instruments, whose low-energy detection threshold was around 10$^{26}$\,Hz. It suggested that the brightest $\gamma$-ray objects had the bulk of their emission at frequencies well below the VHE $\gamma$-ray range. 

However, this sequence was built with just a hundred sources and, in case of the $\gamma$-ray band, only $\sim$30 sources were adopted to build the average SEDs.
Moreover, the average SEDs were built not taking into account the simultaneity of the data in different bands. While the former criticism was overcome a few years ago, \S~\ref{sec:fermiseq}, the latter is still a matter of debate, \S~\ref{sec:six}.

\begin{figure}
    \centering
    \includegraphics[width=0.45\textwidth]{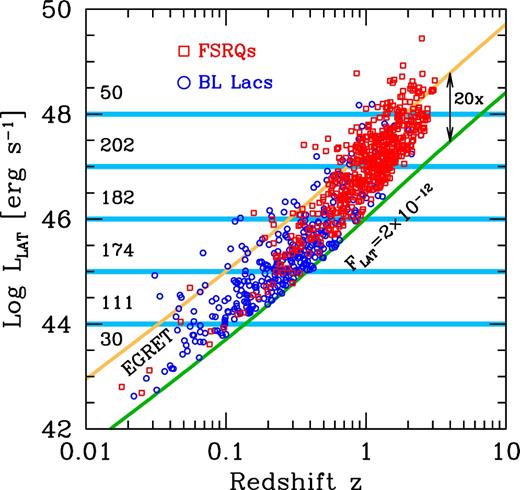}
    \caption{Luminosity in the 0.1--100\,GeV band as a function of redshift for  the \textit{Fermi} sequence sample, from \citet{2017MNRAS.469..255G}. The solid lines represent \egret~ sensitivity limit and \textit{Fermi}-LAT sensitivity after 4 years of observations. Horizontal lines delimit the luminosity bins adopted in the study.}
    \label{fig:fermi_seq_L}
\end{figure}

\subsection{The \textit{Fermi} blazar sequence}\label{sec:fermiseq}

Almost 20 years after the publication of the blazar sequence paper and motivated by the detection, in the $\gamma$-ray band, of about 1500 blazars by the \textit{Fermi}-LAT satellite as reported in the 3LAC catalog \citep{2015ApJ...810...14A}, \citet{2017MNRAS.469..255G} proposed a revised sequence, named \textit{Fermi} sequence. 
This time, the authors explored the existence of a SED sequence on a sample of $\gamma$-ray selected blazars binned according to the luminosity in the MeV-GeV energy range. 

The source sample was extracted from the 3LAC catalog, a complete, flux-limited sample built with 4 years of \textit{Fermi}-LAT observations. The full catalog includes 1563 sources identified with AGNs at high galactic latitudes, with 98\% blazar associations. After removing sources with no clear association, no redshift determination,  and/or not associated with blazars, the authors built a sample of 747 sources (299 classified as BL Lacs and 448 as FSRQs).  

The 0.1--100\,GeV band K-corrected luminosity as a function of redshift is represented in Fig. \ref{fig:fermi_seq_L}. 
The redshift distribution reaches $z \sim 3$ and the luminosity range spans several orders of magnitude: 
from $10^{42}$ to almost $10^{50}$\,erg/s. In general, BL Lac objects are located at smaller 
$z$, mostly below $z \sim 1$, and exhibit lower average luminosity. FSRQs are instead located 
at intermediate and large $z$ and exhibit larger average luminosity.

For the construction of the revised sequence, the sample was divided into six luminosity bins, delimited by horizontal lines in the figure. In analogy with the original sequence, each luminosity bin spans a decade in gamma-ray luminosity. This allows for an easy comparison between the two sequences.
Table~\ref{tab:fermi_seq} lists the luminosity bins adopted in the study. In the table, the main phenomenological parameters describing the average SED are reported, namely, the synchrotron peak position ($\nu_S$), the inverse Compton peak position ($\nu_{IC}$), and the Compton dominance value (CD). 

\begin{figure}
 \vskip -2 cm
   \centering
 \includegraphics[width=0.6\textwidth]{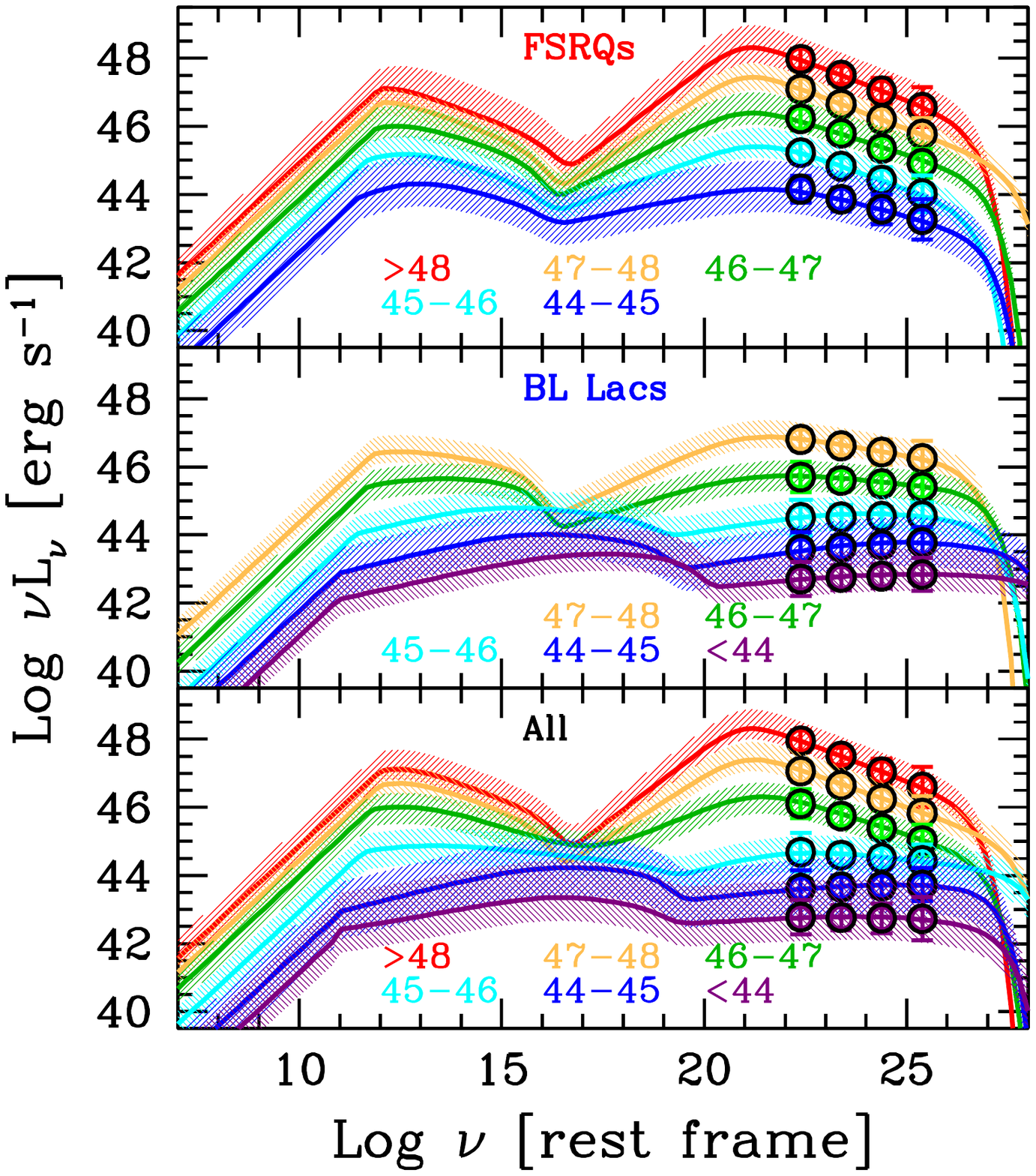}
\vskip -2 cm
    \caption{The {\it Fermi} blazar sequence, from \citet{2017MNRAS.469..255G}. FSRQs (top panel) of different luminosity classes do show an increasing  Compton dominance for increasing luminosity, but not the trend of decreasing peak frequencies. 
    This behaviour, instead is clearly evident when considering BL Lac objects (mid panel) and also when considering
    all blazars, without dividing them into FSRQs and BL Lacs.}
    \label{fig:fermi_seq}
\end{figure}

The last three columns of the table report the total number of objects in each bin, the fraction of FSRQs, and that of BL~Lac objects, respectively. Above $10^{45}$\,erg/s the sample is dominated by FSRQs, while at low luminosity BL Lac objects are dominant. Only 2 BL Lac objects display an average luminosity >$10^{48}$\,erg/s. On the other hand, 9 FSRQs belong to the <$10^{44}$\,erg/s luminosity bin, as can be seen in Fig~\ref{fig:fermi_seq_L}. 
The {\it Fermi} sequence is illustrated in Fig.~\ref{fig:fermi_seq}: in 
the bottom panel FSRQs and BL Lacs are shown together, while 
in the top and central panel they are shown separately.
This distinction allows for a comparison of the two source classes.

When divided into luminosity bins, the \textit{Fermi}-selected blazars form a sequence with the same average properties of the original sequence. We notice a shift in both synchrotron and inverse Compton peak position with decreasing luminosity. A change in CD is also evident. When considering BL~Lac objects and FSRQs separately, however, some additional features emerged in the study. The peak shift is mostly affecting BL Lac objects and not FSRQs. On the other hand,  the increase in CD with increasing luminosity is only evident in the FSRQ subsample.

The more complex scenario emerged in this study with respect to the original one might be in part explained by the presence, in this new sequence, of blazars with a wide range of black hole masses. While very likely the original sequence concerned only BL~Lacs and FSRQs with black holes with large masses, the improved sensitivity at all frequencies allowed to detect also objects of smaller masses. An effect evident in particular in the low-luminosity bins, where nearby intermediate mass FSRQs coexist with nearby BL Lacs with higher masses. The net result is a large dispersion of the points in the broadband SEDs of Fig.~\ref{fig:fermi_seq}. 

\begin{table}[]
    \centering
    \begin{tabular}{c|c|c|c|c|c|c}
    \hline
    \hline
    log~$L_\gamma$ [log(erg/s)] & $\nu_S$ [Hz]  & $\nu_{IC}$ [Hz] & CD & N$_{All}$ & N$_{FSRQ}$ & N$_{BL\,Lac}$ \\
    \hline
    \hline  
      >48    & 2.5e12 & 9e20 & 15  & 49 & 47 (96\%) & 2  \\
      47--48 & 2.5e12 & 2e21 & 4.8 & 202 & 177 (88\%) & 25 \\
      46--47 & 5e12 & 2e21 & 2    & 182 & 144 (79\%) & 38 \\
      45--46 & 5e12 & 1e22 & 0.6  & 174 & 52 (30\%) & 122 \\
      44--45 & 1e16 & 8e24 & 0.35 & 111 & 19 (17\%) & 92 \\
      <44    & 4e16 & 3e25 & 0.25 & 29  & 9 (31\%) & 20 \\
    \hline
    \hline
    \end{tabular}
    \caption{Main parameters characterizing the \textit{Fermi}~blazar sequence \citep{2017MNRAS.469..255G}.}
    \label{tab:fermi_seq}
\end{table}

\subsection{The TeV blazar sequence}\label{sec:tevseq}

In recent years, the rapid development of Cherenkov telescopes allowed the detection of  a significant number of blazars in the TeV energy range. 
In the TeVCat catalogue\footnote{\url{http://tevcat2.uchicago.edu/}} there are now 81 blazars, and 67 of these have a measured redshift.
Using the SSDC Sky Explorer facility \footnote{\url{https://tools.ssdc.asi.it/}} we have constructed the SED for each of these sources. 
We have then divided FSRQs and BL Lacs, and using SSDC data we have divided the latter into the same $\gamma$-ray luminosity bins adopted for the {\it Fermi} blazar sequence, to allow for a direct comparison. 

\begin{figure}
\vskip -2cm`
\centering
    \includegraphics[width=0.6\textwidth]{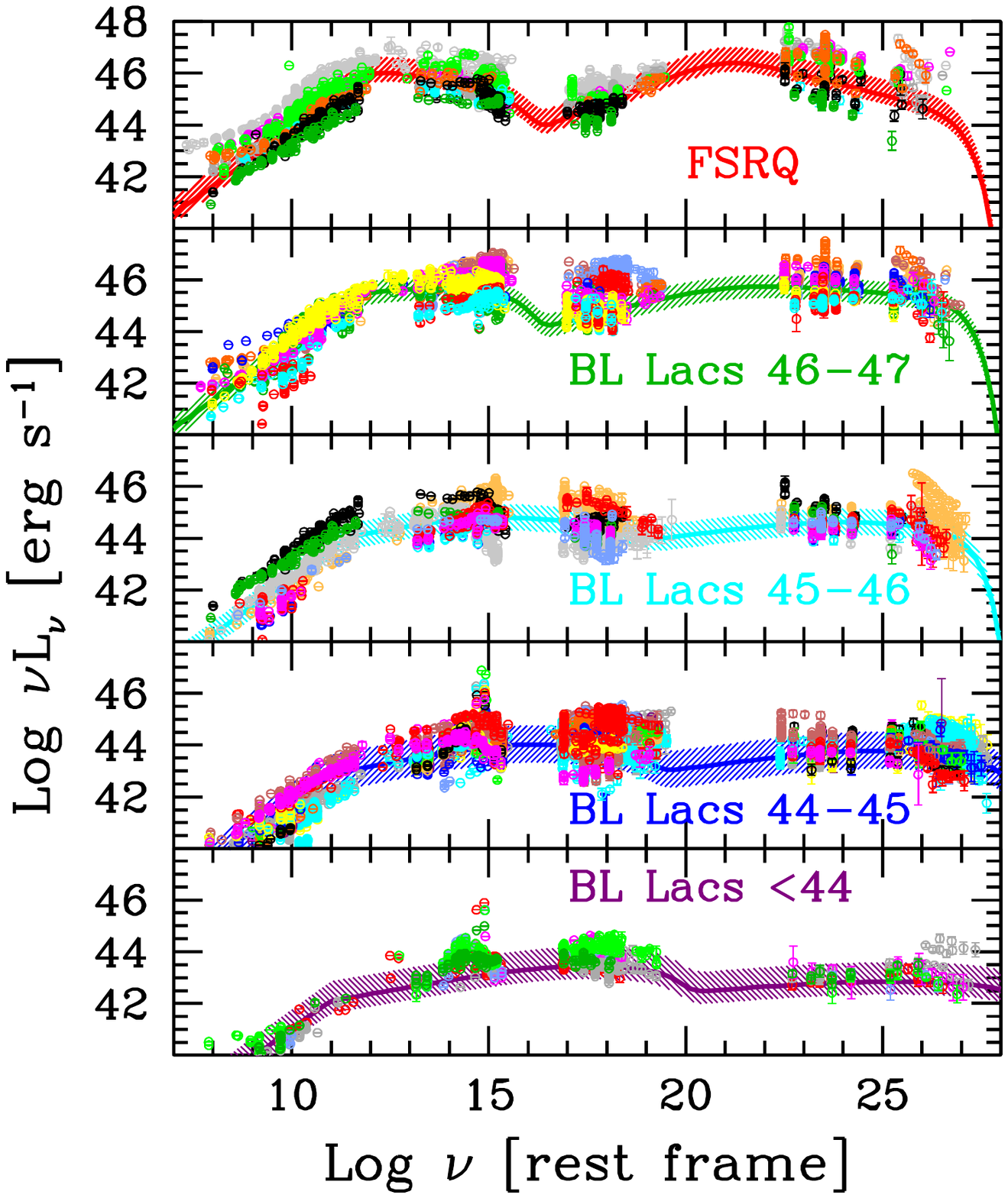}
\vskip -2 cm
    \caption{The TeV blazar sequence. The SED of all TeV detected blazars with known redshift have been collected and divided into the luminosity bins adopted in the {\it Fermi} blazar sequence \citep{2017MNRAS.469..255G}. The stripes are the 1$\sigma$ dispersion from \citep{2017MNRAS.469..255G}.}
    \label{fig:TeV_seq}
\end{figure}

The result is shown in Fig.~\ref{fig:TeV_seq}, where the source data are compared with the 1-$\sigma$ stripes corresponding to the average SED derived in \citet{2017MNRAS.469..255G} for the same luminosity bin. 
We can see that there are no strong differences between the SED of the blazars detected by \textit{Fermi} and the ones that have also been detected in the TeV band.
However, there is indeed an indication, for the brightest TeV BL Lacs, that their X-ray luminosity is on average larger than the average of the \textit{Fermi} blazars with the same luminosity in the \textit{Fermi} energy range.

A possible explanation of this effect is that some of the blazars in this luminosity bin are blue quasars. 
These sources would not contradict the proposed explanation of the sequence based on radiative cooling (\S\ref{sec:five}), because they could be FSRQs that produce most of their power beyond the typical distance of the BLR. 
The deficit of seed photons for scattering in these sources implies reduced radiative losses and hence a large $\gamma_{\rm peak}$, resulting in a blue spectrum \citep{2008MNRAS.387.1669G}. 
Alternatively, this could be simply due to a selection effect and they might represent the few bright existing HBLs. Since the TeV sample is strongly biased toward sources with high X-ray luminosity (one of the criteria used for the identification of the best targets), it is natural that the first sources detected in these energy bins are the brightest in the X-ray range. 

A third plausible option is that these are HBLs with a wrong (too large) redshift determination, hence their bolometric luminosity is overestimated. Indeed, many of the redshifts determined for HBLs are debated due to the difficulty to detect the host galaxy \citep{2017ApJ...837..144P}. We will investigate this issue in a future paper (Prandini et al. in prep.).

\section{At the extremes of the blazar sequence}\label{sec:four}
The most recent sequence proposed in \citet{2017MNRAS.469..255G} includes 747 sources, a sample almost six times larger than the original one used in \citet{1998MNRAS.299..433F}. 
Despite the large number of sources considered, however, two subclasses of sources might still elude the study. These are the MeV blazars and the extreme-TeV blazars. 
The former have a powerful high energy peak up to a few MeV, and therefore could remain undetected by the \textit{Fermi}-LAT that observes in the energy range beyond 100\,MeV. 
Extreme-TeV blazars instead peak at a few TeV and are expected to have a steady but faint emission in the GeV-TeV range. Thus, the fainter or more distant objects might easily remain below the sensitivity limit of \textit{Fermi}. These subclasses represent the two extremes of the sequence and their detection by future experiments in the MeV and TeV range is of great importance for testing the current undestanding of the blazar sequence. 

\begin{figure}
    \centering
    \includegraphics[width=0.32\textwidth]{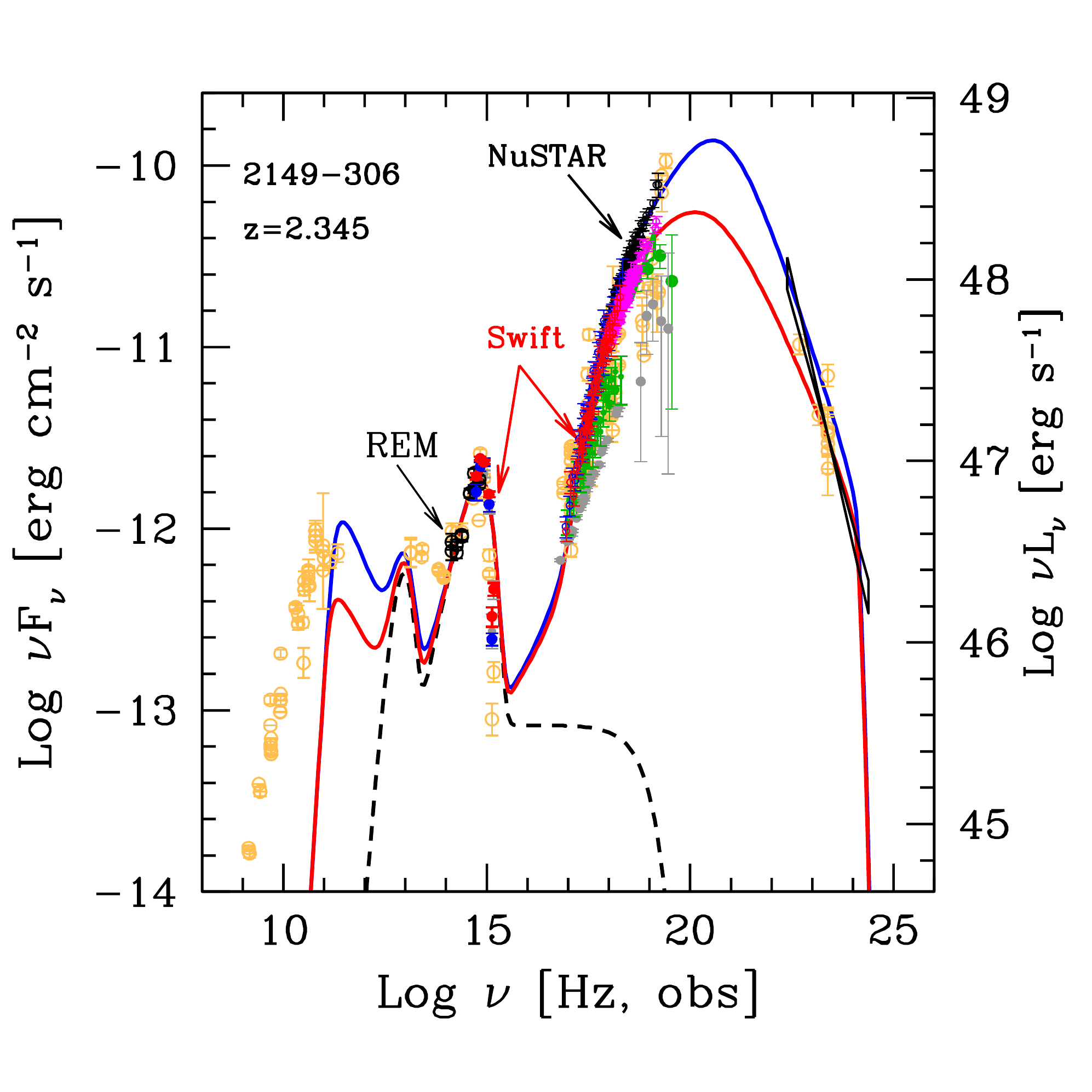}
    \includegraphics[width=0.32\textwidth]{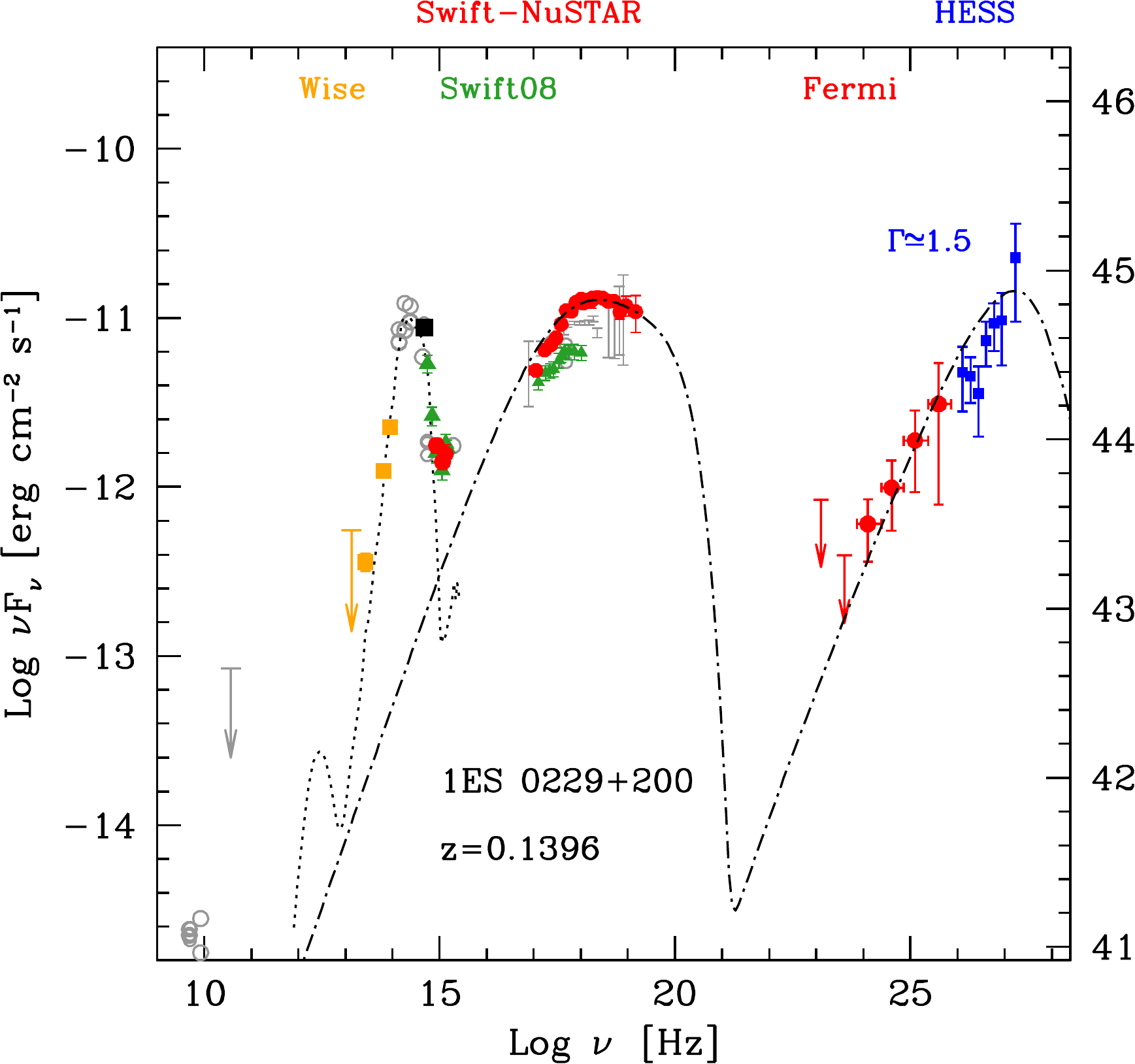}
    \caption{Extremes of the blazar sequence:  SED of a MeV blazar (left) and of a TeV blazar (right). 
    The MeV blazar is the FSRQ 2149--306, located at $z=2.354$, from \citet{2015ApJ...807..167T}, \copyright AAS. Reproduced with permission. 
    The TeV blazar is, instead, 1ES~0229+200, at $z=0.14$. Adapted from \citet{2020NatAs...4..124B}.} 
    \label{fig:MeV-blazar}
\end{figure}

\subsection{Extreme TeV blazars}

Out of the several tens of VHE gamma-ray emitting blazars discovered so far, roughly 1/4 are classified as extreme blazars \citep[see][and references therein]{2020NatAs...4..124B}. 
These extreme blazars come in two flavours: extreme synchrotron and extreme TeV sources. 
The former are characterized by a synchrotron peak beyond 1\,keV, while the latter exhibit a $\gamma$-ray peak energy above 1\,TeV. 
The prototypical extreme TeV blazar is 1ES~0229+200, Fig.~\ref{fig:MeV-blazar} (right). 
Located at redshift 0.14, this blazar holds the record for the highest 
$\gamma$-ray peak frequency, exceeding 10\,TeV. 
As reported in \citet{2020NatAs...4..124B}, all 14 known extreme TeV blazars but one 
are also extreme synchrotron sources. 
This suggests that reaching $>$TeV peak energy requires also $>$keV synchrotron peak frequencies, 
hence an efficient electron acceleration up to high energies. 
Another key characteristic of this class of sources is their faint luminosity. 
For the known extreme TeV blazars, the luminosity measured barely exceeds 
the \textit{Fermi}-LAT sensitivity limit in the gamma-ray range. 
This might suggest that the extreme TeV blazars identified so far are only the tip of the (faint) iceberg of the extreme TeV blazar population. 
The bias is due to the selection effect in the VHE energy range, where the observations in the last 15 years have been driven by the \textit{Fermi}-LAT properties reported in the published catalogs.

In the context of the \textit{Fermi} blazar sequence, these still missing extreme TeV blazars would 
populate the lowest luminosity bin. Future gamma-ray instruments like LHAASO \citep{2019arXiv190502773B}, SWGO \citep{2019arXiv190208429A}, ASTRI \citep{2020JHEAp..26...83P}, and CTA \citep{2019scta.book.....C}, will be the key to probe this expected trend. Wide field-of view instruments, like LHAASO and SWGO, will have the advantage of a wide field of view, but a limited sensitivity at sub-TeV energies, limiting the survey to nearby objects ($z\leq0.2$).
Future IACTs, instead, will access a parameter space (GeV-TeV energy, meaning large distance $z>1$, and low luminosity) still not explored by the current generation of gamma-ray instruments 
 with a limited field of view. Therefore, targeted observations on promising objects \citep[e.g.,][]{2020MNRAS.491.2771C} will be performed in this case.


\subsection{MeV blazars}

If we now focus the attention on the opposite side of the sequence, we might wonder if there are missing objects peaking at low frequencies. 
Following the sequence main trend, these would be most likely very powerful FSRQs peaking in the $\sim$meV and MeV energies, respectively. 
The \textit{Fermi}-LAT non-detection is due to the fact that the high energy emission does not reach the LAT energy band. 

Interestingly, a number of recent studies in the hard X-ray band have consolidated the 
existence of ultra-luminous FSRQs, located at a large redshift ($z>2$). 
In addition to the large distance, these objects are characterized by an extremely massive 
central black hole, estimated to be among the largest known \citep[see e.g.,][]{2021arXiv210908156S}. 
The main factor limiting \textit{Fermi}-LAT detection of this subclass of objects is their 
distance, that affects the observed flux. 
An example of a powerful MeV blazar is 2149--306 whose SED is shown in Fig. \ref{fig:MeV-blazar} (left).
The source is located at redshift 2.354.  
\citet{2018JHEAp..19....1D} show that if this blazar were located at
$z=7$, it would be out of reach for current generation of $\gamma$-ray telescopes. 
Still, it would be detectable in the MeV range by future generations of MeV satellites such as 
$eASTROGAM$ \citep{2018JHEAp..19....1D}, $AMEGO$ \citep{2019BAAS...51g.245M}, or COSI \citep{2019BAAS...51g..98T}.

\section{The blazar sequence: theoretical approach}\label{sec:five}

The original blazar sequence was soon \citep{1998MNRAS.301..451G} interpreted as a sequence of radiative cooling. 
The main idea is that the emitting electrons cannot achieve high energies if their radiative cooling
(by the synchrotron, Self Compton and External Compton processes) is severe. 
According to this idea, in FSRQs the broad line region and the molecular 
torus are essential sources of seed photons for the External Compton 
process, and this implies that the random Lorentz factor of the electrons 
emitting at the peaks of the SED, $\gamma_{\rm peak}$, is relatively small.
Therefore, the peak of their synchrotron emission occurs in the sub-mm band, 
while their high energy peak is in the MeV band. 
Another important consequence is that the inverse Compton process becomes 
dominant with respect to the synchrotron one, implying a large Compton dominance.
In BL Lacs, the absence of an important thermal source implies the paucity 
of seed photons for Compton scattering. 
The radiative cooling is less severe, allowing electrons to achieve large $\gamma_{\rm peak}$. 
Thus, the synchrotron peaks in the UV--X--ray band, and the high energy 
hump peaks in the GeV and sometimes even in the TeV band.
The lack of ambient seed photons implies that only the SSC process contributes 
to the high energy emission.
As a result, in BL Lacs the high energy component is usually not dominant.

\begin{figure}
    \centering
    \vskip -2 cm
    \includegraphics[width=0.4\textwidth]{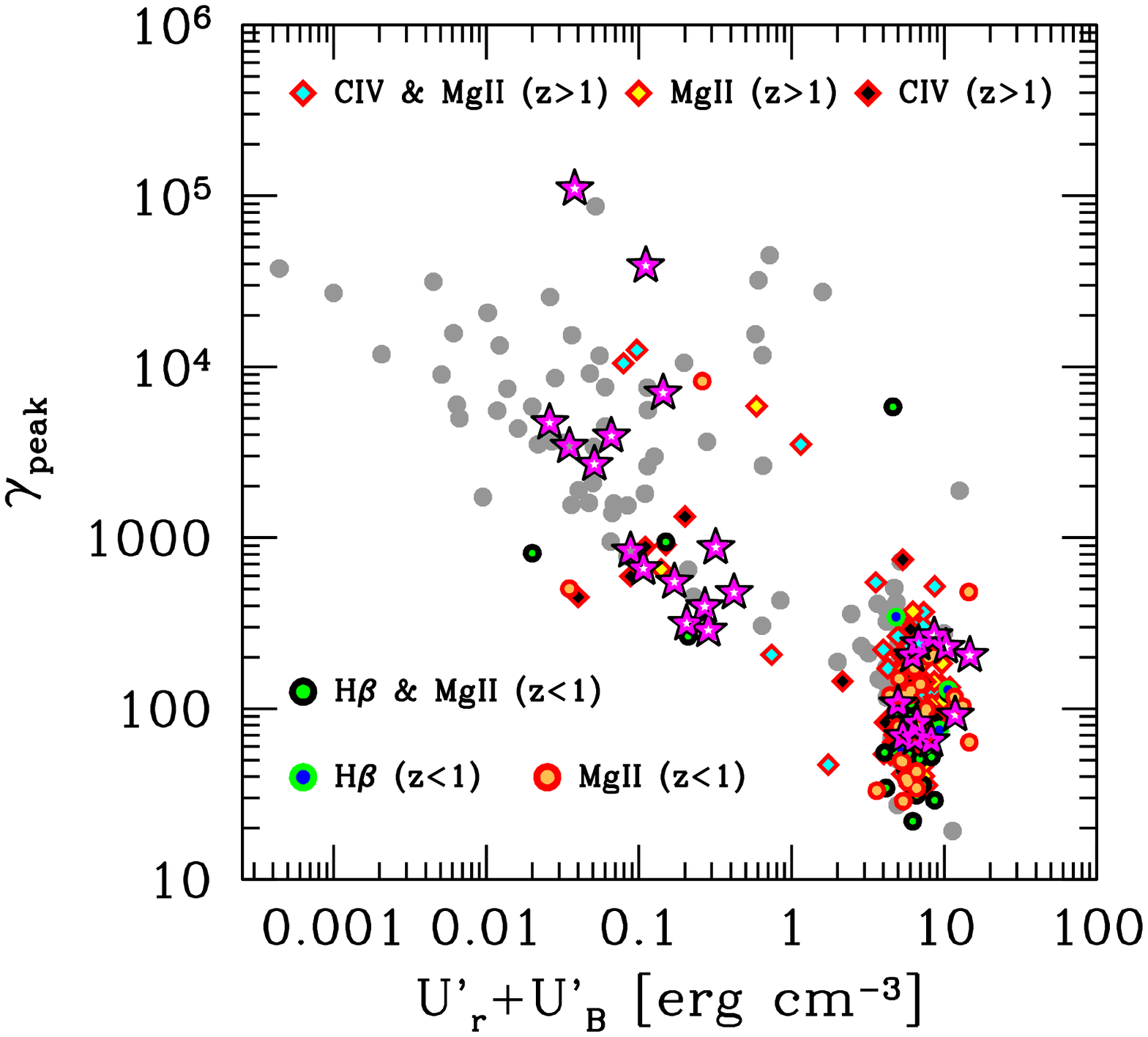}
    \vskip -1.5 cm
    \caption{The energy $\gamma_{\rm peak}$ 
    of the electrons emitting at the peaks of the SED as a function of the energy density (magnetic plus radiative) as seen in the comoving frame. 
    Grey symbols refer to blazars studied previously in \citet{2008MNRAS.385..283C,2010MNRAS.402..497G}. The stars are objects with weak and very weak broad emission lines that can be classified as BL Lacs or transitional objects. Adapted from \citet{2015MNRAS.448.1060G}. 
}
    \label{gpeak_u}
\end{figure}

\subsection{Physical parameters}

This idea can be checked by applying a leptonic one-zone model, 
that allows to infer the physical parameters of
the different classes of sources. 
Fig. \ref{gpeak_u} shows the random Lorentz factor $\gamma_{\rm peak}$ 
of the electrons contributing to the peaks of the SED as a function of the 
total (magnetic plus radiative) energy density as measured in the comoving 
frame  \citep{2015MNRAS.448.1060G,2008MNRAS.385..283C, 2010MNRAS.402..497G}. There is a clear trend: the largest electron energies are possible only 
for small values of $U^\prime_B+ U^\prime {\rm _r}$.

We also know that the presence or absence of broad emitting lines is associated with 
the luminosity of the accretion disk and its accretion regime.  
Furthermore, there is a relation between the jet power and the disk luminosity.
Therefore, the blazar sequence can be thought also as a sequence whose
main parameter is the jet power or, equivalently, the disk luminosity. 
On the other hand, we must consider also the mass of the central black hole, 
since the accretion regime changes when the disk luminosity is of the order 
of $10^{-3}$--$10^{-2}$ of the Eddington one.

The fact that, for FSRQs, there is no trend between luminosity and 
location of the peak frequencies can be readily explained if the emission 
occurs within the distance of influence of the BLR and/or the molecular torus. 
In fact, the BLR radius is proportional to $L_{\rm disk}^{1/2}$: this 
implies that the radiation energy density of the line photons ($U_{\rm BLR} 
\sim 0.1 L_{\rm disk}/(4\pi R^2_{\rm BLR} c$) is constant as long as the emission 
site is within $R_{\rm BLR}$: $U_{\rm BLR}\sim 1/(12\pi)$.
A similar conclusion holds also for the energy density of the torus 
radiation, once the torus is approximated as a portion of a sphere of 
radius $R_{\rm T}$ intercepting and re-emitting a fraction $f$ of 
$L_{\rm disk}$: $U_{\rm T}\sim 0.07 f/(12\pi)$ \citep{2015MNRAS.448.1060G}.
Both $U_{\rm BLR}$ and $U_{\rm T}$ are seen boosted by a $\Gamma^2$  factor in the comoving frame.

 We note that \citet{2018MNRAS.477.4749C} argued that
the fingerprint of the $\gamma$--$\gamma$ $\to$ $e^+ e^-$ 
process due to the presence of emission line photons
is not visible in the {\it Fermi}-LAT spectra of blazars  \citep[but see][]{2010ApJ...717L.118P}. To the aim of explaining the sequence in terms of different radiative cooling, 
the fact that the location of the emitting region is inside the BLR sphere
or outside it, but within the torus, is not very important. 
Within the BLR the corresponding radiation energy density is larger,
and timescales are shorter. Within the torus the external radiation energy density
is smaller, but the size of the source is probably larger, so that the
ratio between the dynamical and the cooling timescales in the two zones are 
not very different.

For BL Lacs, the absence of these structures implies that the radiation 
energy density is of synchrotron origin only, therefore associated to 
the jet power and luminosity, and we do see a trend of larger $\gamma_{\rm peak}$ 
for smaller $(U^\prime_B+ U^\prime {\rm _r)}$. 

\section{Criticisms to the blazar sequence}\label{sec:six}

Since its birth in 1998, the blazar sequence was and is a matter of active debate. 
We can divide the main criticisms roughly into two categories: in the first the validity of the observed sequence is accepted, but the interpretation
is different from the original one (that assumes that what drives the sequence is the difference in radiative cooling); for the second category, instead, 
the sequence is not real, it is only the results of observational biases.

Into the first category there is the paper by \citet{2017ApJ...835L..38F}, extending the work of \citet{2008A&A...488..867N}. They explain the observed blazar sequence as a sequence of Doppler factors, 
that boosts the intrinsic luminosity and the intrinsic peak frequencies.  They claim that in the comoving frame the sequence should have just the opposite trend.
Consider that the method to derive the beaming factor uses the brightness temperature compared with the equipartition value of $5\times 10^{10}$~K
\citep{1994ApJ...426...51R}, and the  minimum variability timescale observed in the radio. 
This implies that sources that are not varying during the observing periods are bound to have very small Doppler factors, and in fact they range from $\sim 1$ to $\sim 30$. It is thus possible that some sources are under-corrected. Consider also that the beaming in the radio band may be different from the beaming
of more compact and innermost regions producing the bulk of the emission.

Alternatively, the sequence has been interpreted by \citet{2015MNRAS.453.4070P}, 
in their inhomogeneous jet model, as a consequence of the different location of the dissipation region: in BL Lacs is closer to the central engine than in FSRQs, resulting in stronger magnetic fields, smaller Compton dominance, and larger peak frequencies with respect to FSRQs.

In the second category  we find, among others, the study of \citet{2021MNRAS.505.4726K}, 
considering more than a thousand blazars of both the FSRQ and BL Lac type.
They conclude that FSRQs, while spanning a large range of observed luminosity, are all low peaked blazars.
More importantly, there is no luminosity - peak frequency trend in BL Lacs. In their Fig. 4  that illustrates this point, one notices indeed a very large spread of peak frequency for any luminosity for BL Lacs, but there is still a trend of larger 
synchrotron peak for smaller luminosities. 

A detailed proposal questioning the reality of the sequence was put forward by \citet{2012MNRAS.420.2899G} proposing what they call a `simplified scenario’ for blazars.
According to these authors, the blazar sequence is a result of selection effects with no physical link between the luminosity and the overall blazar SED.
Both at high and low luminosities, we have both red and blue blazars. 
In this scenario, the distribution of $\gamma_{\rm peak }$ is assumed to peak at $\sim 10^3$, and it is asymmetric, with a high-energy tail broader than the low-energy one (see Fig. 4 of \citep{2012MNRAS.420.2899G}). 
Furthermore, the radio luminosity function is $N(L) \propto L^{-3}$, the magnetic field is assumed to be $B=0.15$~G for all blazars, and the mean Doppler factor distribution is $<\delta> =15\pm 2$. 
The immediate consequence is that there exist many BL Lacs of low  power with a ``red" SED. 
These are the ones with $\gamma_{\rm peak}\sim 10^3$ or smaller. 
We do not see them yet because, being at low luminosity, we do not have a survey deep enough to let them emerge. 
These objects should have a high energy, synchrotron self-Compton spectrum peaking in the 1–10 keV band (see Fig. 11 of \citep{2017MNRAS.469..255G} for illustration of this point).
According to this idea, there should be BL Lacs already detected in the radio, 
at moderate (but still unknown) redshifts, with no X-ray detection yet. On the other hand, with the advent of $eROSITA$ \citep{2012arXiv1209.3114M}, and further in time $Athena$ these objects would be detected in X-rays, helping to probe their BL Lac nature.


\section{Future perspectives}\label{sec:seven}
After more than 20 years from its proposition, the blazar sequence is still alive and lively discussed. It has been deeply investigated from both the experimental and theoretical point of view. 
The original trend is confirmed by recent, extensive data in $\gamma$ rays, even if BL~Lac objects and FSRQs seem to behave differently, but consistently with the theoretical expectation. The emerging sample of TeV-detected blazars, despite a strong selection bias, offers a new interesting perspective that will be further investigated. 

Interestingly, \citet{2018ApJ...854...54R} studied the neutrino and ultra-high energy cosmic ray (UHECR) emission based on the model \citep{2014PhRvD..90b3007M} in the context of the blazar sequence. A strong connection is reported, with HBLs tending to produce UHECRs, while FSRQs are more efficient neutrino emitters. 
In \citep{2009ApJ...702..523I} the blazar sequence is used to estimate the 
gamma-ray background. 

From the theoretical side, the sequence has been interpreted as an effect of the efficiency of the cooling mechanism on accelerated electrons. In FSRQs, the cooling is efficient, hence preventing electrons to reach high energies and causing a large Compton dominance. In BL Lacs, instead, the cooling is less effective allowing electrons to reach the highest energies. Remarkably, this efficiency turned out to be intimately connected with the accretion regime. In this perspective, the blazar sequence is the result of a change in the accretion regime: efficient accretion (high disk luminosity) is the cause of efficient cooling. Alternative theoretical interpretations have been proposed, invoking changes in the Doppler factor \citep{2008A&A...488..867N,2017ApJ...835L..38F} or in the location of the dissipation region \citep{2015MNRAS.453.4070P} as possible explanations to the observed sequence.

The very existence of the sequence is the target of criticisms, mostly based on selection effect arguments \citep{2012MNRAS.420.2899G,2021MNRAS.505.4726K}.  Recent and near-future observatories represent a unique opportunity to solve the dispute and probe (or reject) the validity of the sequence. Moreover, future observations will contribute to solve the main open points about the sequence, namely: i) evaluate the existence of a population of MeV blazars and, on the opposite side, that of extreme blazars; ii) probe the properties of the accretion disk, in particular in the BL~Lac object case (ADAF?); iii) evaluate the effect of cosmic evolution.

The advent of X-ray and $\gamma$-ray surveys with $eROSITA$, LHAASO, SWGO in the current decade and, later on, that of  $Athena$ \citep{2013arXiv1306.2307N},  will provide an unbiased census of blazars. This will complement very deep observations in other bands (e.g., SKA and JWST \citep{2015aska.confE.174B,2006SSRv..123..485G}). 
Finally, the recently launched $IXPE$ mission \citep{2016SPIE.9905E..17W} is going to characterize for the first time the X-ray polarized emission from some bright selected sources, improving significantly our understanding of the blazar phenomena. 
Despite all these exciting advancements foreseen in the near future, we point out that the lack of \textit{Fermi}-LAT full sky monitoring successor and that of \textit{Swift}-XRT and UVOT fast reactions to transients and ToOs will strongly affect these observations. 

A complementary approach to the study of the sequence that might become accessible thanks to the improved sensitivity of future instruments is the study of misaligned blazars, i.e., radio galaxies. However, the emerging evidence of composite jets in blazars (internal, fast spine, and external, slow layer) represents a non-trivial aspect to be taken into account when trying to connect blazars and radio galaxies.

\vspace{6pt} 

\funding{E.P. acknowledges funding from Italian Ministry of Education, University and Research (MIUR) through the "Dipartimenti di eccellenza” project Science of the Universe.}

\dataavailability{For the study presented in \S~3.3, we have used the following public archives:
\begin{itemize}
\item[-] TeVCAT online catalog for TeV Astronomy, \url{http://tevcat2.uchicago.edu/}
\item[-] SSDC Sky Explorer facility, \url{https://tools.ssdc.asi.it/}
\end{itemize}
}

\acknowledgments{We would like to thank L. Costamante for re-adapting Fig.~\ref{fig:MeV-blazar} for this manuscript. We are grateful to the
reviewers for their constructive feedback on the manuscript.}

\conflictsofinterest{The authors declare no conflict of interest.} 


\reftitle{References}


\end{document}